\documentclass[english,conference]{IEEEtran}
\usepackage[T1]{fontenc}
\usepackage[latin9]{inputenc}
\usepackage{color}
\usepackage{amsthm}
\usepackage{amsmath}
\usepackage{amssymb}
\usepackage{graphicx}

\makeatletter

\providecommand{\tabularnewline}{\\}

\theoremstyle{plain}
\newtheorem{thm}{\protect\theoremname}
\theoremstyle{definition}
\newtheorem{example}[thm]{\protect\examplename}

\ifCLASSINFOpdf
\else
\fi
\makeatother

\usepackage{babel}
\providecommand{\examplename}{Example}
\providecommand{\theoremname}{Theorem}

\begin{document}

\title{Blind Interference Alignment in General Heterogeneous Networks}

\author{\IEEEauthorblockN {Vaia Kalokidou} \IEEEauthorblockA{Communication
Systems and\\Networks Research Group\\MVB, School of Engineering\\University
of Bristol, UK\\ Email: eexvk@bristol.ac.uk}\and\IEEEauthorblockN
{Oliver Johnson} \IEEEauthorblockA{ Department of Mathematics\\University
of Bristol, UK\\ Email: o.johnson@bristol.ac.uk}\and \IEEEauthorblockN{Robert
Piechocki} \IEEEauthorblockA{Communication Systems and\\ Networks
Research Group\\MVB, School of Engineering\\University of Bristol,
UK\\ Email: r.j.piechocki@bristol.ac.uk}}
\maketitle
\begin{abstract}
Heterogeneous networks have a key role in the design of future mobile
communication networks, since the employment of small cells around
a macrocell enhances the network's efficiency and decreases complexity
and power demand. Moreover, research on Blind Interference Alignment
(BIA) has shown that optimal Degrees of Freedom (DoF) can be achieved
in certain network architectures, with no requirement of Channel State
Information (CSI) at the transmitters. Our contribution is a generalised
model of BIA in a heterogeneous network with one macrocell with $K$
users and $K$ femtocells each with one user, by using Kronecker (Tensor)
Product representation. We introduce a solution on how to vary beamforming
vectors under power constraints to maximize the sum rate of the network
and how optimal DoF can be achieved over $K+1$ time slots. 
\end{abstract}

\section{Introduction}


Next-generation mobile and cellular networks will require higher capacity
and reliabilty, as well as power-efficiency. Interference Alignment
(IA), first introduced by Maddah-Ali, Motahari and Khandani in {[}1{]}
and Cadambe and Jafar in {[}2{]}, made a very promising step in this
direction by proving that it is possible that the $K$-user interference
channel, under the assumption of global perfect CSI, can have $K/2$
DoF, i.e. ``everyone gets half the cake''. The novelty of the IA
scheme, as described in {[}1{]}-{[}3{]}, lies in the fact that it
attempts to align, rather than cancel or reduce, interference along
dimensions different from the dimensions of the actual signal. 

Initially, the main drawbacks of IA were the requirement of global
perfect CSI at the transmitter (CSIT), which resulted in feedback
overhead, and its complexity, as only for the $K=3$ case, as presented
in {[}4{]}, a closed-form solution could be easily described. In general,
in the absence of perfect or partial CSIT, the DoF of a network collapse,
i.e transmissions are no longer reliable. However, for certain networks,
the scheme of Blind IA (BIA), originally presented by Wang, Gou and
Jafar in {[}5{]} and Jafar in {[}6{]}, can achieve full DoF, even
when no CSIT is available. BIA can be successfully achieved by a)
knowing distinct coherence patterns associated with different receivers,
or b) employing distinct antenna switching patterns at receivers equipped
with reconfigurable antennas. Furthermore, as suggested by Jafar in
{[}7{]}, BIA can achieve even higher than $K/2$ DoF in certain cellular
environments simply by seeing frequency reuse as a simple form of
IA. Moreover, {[}7{]} introduced the feasibility of BIA in heterogenous
networks due to interference diversity, i.e. the observation that
every receiver experiences a different set of interferers, and depending
on the actions of its interferers, the interference-free signal subspace
fluctuates differently from the rest of the receivers. Finally, {[}8{]}
and {[}9{]} introduced an equal-power allocation BIA scheme that reduces
noise enhancement by constant power transmission. 

In this paper, based on {[}5{]}-{[}7{]}, we propose a generalised
model of BIA in a heterogeneous network, where there is one macrocell
with $K$ users and $K$ femtocells with one user each (see Figure
1). Our contribution is the generalisation of the construction given
by Jafar, {[}7, Section 6{]} in the case $K=2$, introducing the application
of BIA to heterogeneous networks. Moreover, this paper introduces
a new description of the BIA model using a Kronecker Product representation.
Based on our findings, the DoF that can be achieved in both tiers
of the network are presented. Finally, we discuss how to vary parameters
of the model to maximize sum rate, extending the ideas of {[}8{]}-{[}9{]},
and demonstrating optimality in the sum rate sense.

The rest of the paper is organized as follows. Section II presents
the general description of the BIA model, including the determination
of the beamforming matrices, and the whole decoding process. Section
III presents the DoF that can be achieved in the macrocell and the
$K$ femtocells. Section IV presents the achievable sum rate formula
for the heterogenous network. Finally, Section V gives an overview
of our results, illustrated with the aid of simulations/graphs.

\section{System Model}

We generalise Jafar's model {[}7, Section 6{]}, under the same channel
assumptions. Consider the Broadcast Channel (BC) of a heterogeneous
network, as shown in Figure 1, with 1 macrocell and $K$ femtocells.
At the \emph{$N\times N$ }MIMO BC of the\emph{ macrocell}, there
is one transmitter \emph{$T_{xA}$ }with \emph{$N$} antennas, and
\emph{$K$} users equipped with \emph{$N$} antennas each. Transmitter
\emph{$T_{xA}$ }has $N$ messages to send to every user, and furthermore,
when it transmits to user \emph{$a_{k}$}, where $k\in\left\{ 1,2,...,K\right\} $,
it causes interference to all the other $K-1$ users in the macrocell.
At the \emph{$M_{r}\times N$ }MIMO BC of each \emph{femtocell}, there
is one transmitter \emph{$T_{xk}$} with $N$ antennas, and one user
$f_{k}$ equipped with \textbf{$M_{r}$} antennas, with\textbf{ $M{}_{r}=N-1$}.
Transmitter \emph{$T_{xk}$} has \emph{$\mathcal{M}=(T-1)M_{r}+1$}
messages to send to the femtocell user $f_{k}$, and when it transmits
to $f_{k}$, it causes interference to the macrocell user\emph{ $a_{k}$}.
The operation is performed over $T=K+1$ channel uses (i.e. time slots),
which constitute a supersymbol. The channel is assumed to remain constant
over the supersymbol.

The BIA scheme works by using different antenna switching patterns
for each of the $K$ femtocells. These switching patterns are encoded
in the indicator vectors as described later in this section. For the
successful application of the BIA scheme, the following assumptions,
as in {[}7{]}, are made:
\begin{itemize}
\item Users in the femtocells do not receive any interference from transmissions
in the macrocell
\item No CSIT is required, only knowledge of the connectivity of the network
is available at the transmitters
\end{itemize}

\subsection{Beamforming Matrices}

\subsubsection{Macrocell}

The \emph{$((NT)\times1$)} signal at receiver $a_{k}$, for the supersymbol,
is given by:
\begin{equation}
\mathbf{y}^{[a_{k}]}=\mathbf{H}^{[a_{k}]}\mathbf{X}{}_{A}+\mathbf{H}^{[f_{k}a_{k}]}\mathbf{X}_{f_{k}}+\mathbf{Z}^{[a_{k}]}
\end{equation}

Channel transfer matrices are statistically independent due to users'
different locations, and each one of their entries follows an i.i.d.
Gaussian distribution $\mathcal{CN}(0,1)$. $\mathbf{H}^{[a_{k}]}\in\mathcal{\mathcal{C}}^{NT\times NT}$
is the channel transfer matrix from \emph{$T_{xA}$ }to $a_{k}$,
and is given by $\mathbf{H}^{[a_{k}]}=\mathbf{I}_{T}\otimes\mathbf{h}^{[a_{k}]}$
(here and throughout $\otimes$ represents the Kronecker (Tensor)
product), as the channel is non-varying, where $\mathbf{h}^{[a_{k}]}\in\mathcal{\mathcal{C}}^{N\times N}$
is the channel for one time slot. $\mathbf{H}^{[f_{k}a_{k}]}\in\mathcal{C}^{NT\times NT}$
is the inter-cell interference channel transfer matrix from\emph{
$T_{x_{k}}$ }to $a_{k}$, and is given by $\mathbf{H}^{[f_{k}a_{k}]}=\mathbf{I}_{T}\otimes\mathbf{h}^{[f_{k}a_{k}]}$,
where $\mathbf{h}^{[f_{k}a_{k}]}\in\mathcal{\mathcal{C}}^{N\times N}$
is the channel for one time slot. Finally, $\mathbf{Z}^{[a_{k}]}\sim\mathit{\mathcal{CN}}(0,\sigma_{n}^{2}\mathbf{I}_{NT})$
denotes the independent Additive White Gaussian Noise (AWGN) vector.

The ($N\times1$) data stream vector of each user\emph{ $a_{k}$}
is given by $\mathbf{U}^{[a_{k}]}$. The choice of the ($(NT)\times N$)
beamforming matrices \textbf{$\mathbf{V}^{[a_{k}]}$} carrying messages
to users in the macrocell is not unique and should lie in a space
that is orthogonal to the channels of the other $K-1$ macrocell users.
\begin{equation}
\mathbf{V}^{[a_{k}]}=\frac{a}{\sqrt{N}}(\mathbf{v}^{[a_{k}]}\otimes\mathbf{I}_{N}),
\end{equation}
where $a\in\mathrm{\mathfrak{\boldsymbol{\mathfrak{\mathcal{\mathbb{R}}}}}}$
is a constant determined by power considerations (see (4)), and ($T\times1$)
$\mathbf{v}^{[a_{k}]}$ should be a unit vector with entries equal
to $c$, $\sqrt{1-c^{2}}$ (for $c\in\mathrm{\mathfrak{\boldsymbol{\mathfrak{\mathcal{\mathbb{R}}}}}}$
and $c\neq0,\pm1$) or $0$, with a different combination for every
$a_{k}$. For every macrocell user, there will be one time slot in
which only they will be receiving messages. Also, there will be another
time slot (time slot 2 in Figure 2) over which \emph{$T_{xA}$} will
transmit to all users. The\emph{ $((NT)\times1)$} vector $\mathbf{X}{}_{A}$
transmitted by \emph{$T_{xA}$}, is given by:
\begin{equation}
\mathbf{X}{}_{A}=\sum_{i=1}^{K}\mathbf{V}^{[a_{i}]}\mathbf{U}^{[a_{i}]}
\end{equation}

The total transmit power is given by the power constraint: 
\begin{equation}
P_{\mathrm{macrocell}}=\mathbb{E}[\mathrm{tr}(\mathbf{X}_{A}\mathbf{X}_{A}^{T})]=KNa^{2}
\end{equation}

\begin{figure}
\begin{centering}
\includegraphics[width=0.85\columnwidth]{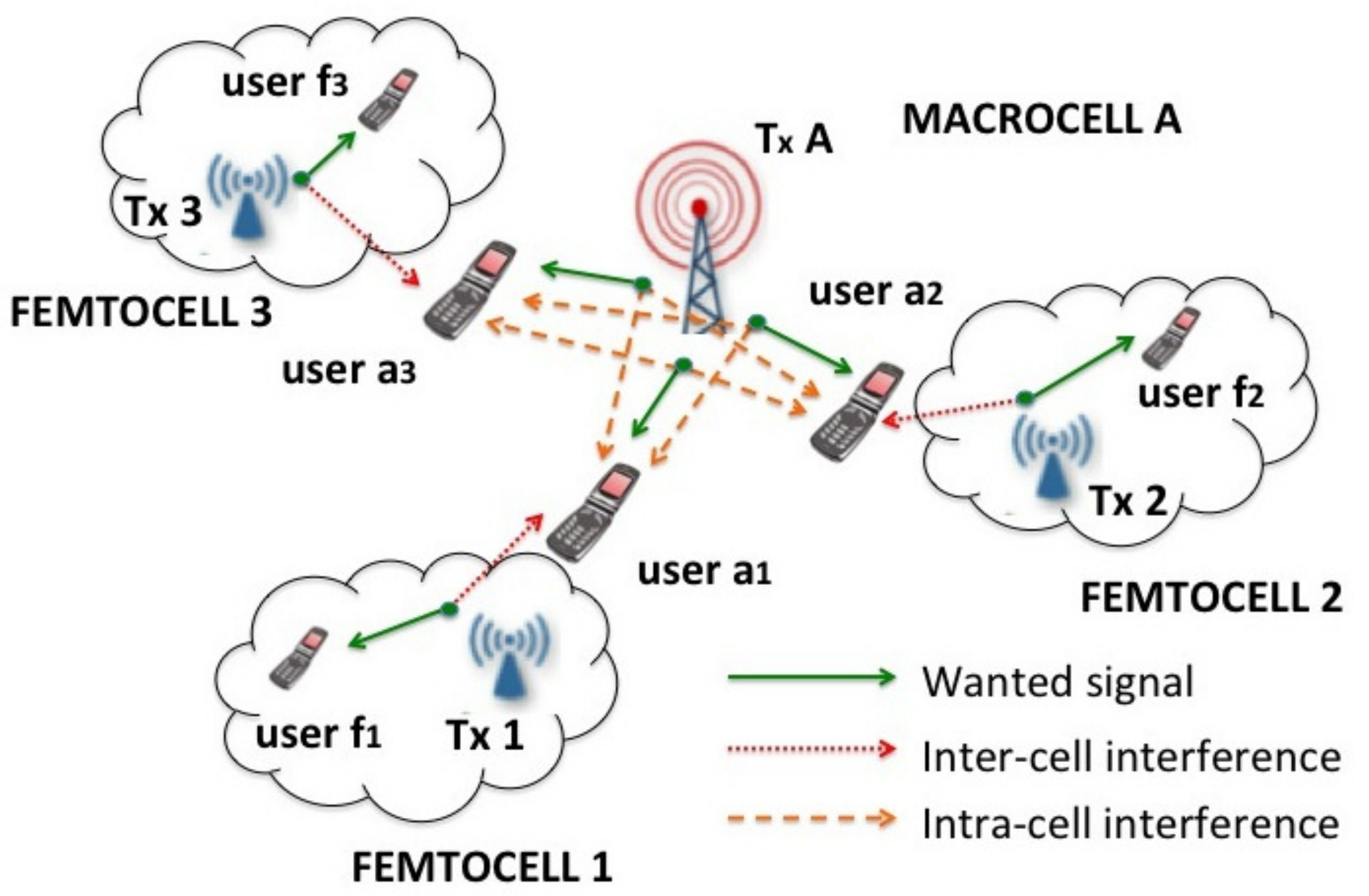}
\par\end{centering}

\caption{BIA in a heterogeneous network: $K=3$ users in the macrocell and
$K=3$ femtocells with 1 user each}
\end{figure}

\begin{figure}
\begin{centering}
\includegraphics[width=1\columnwidth]{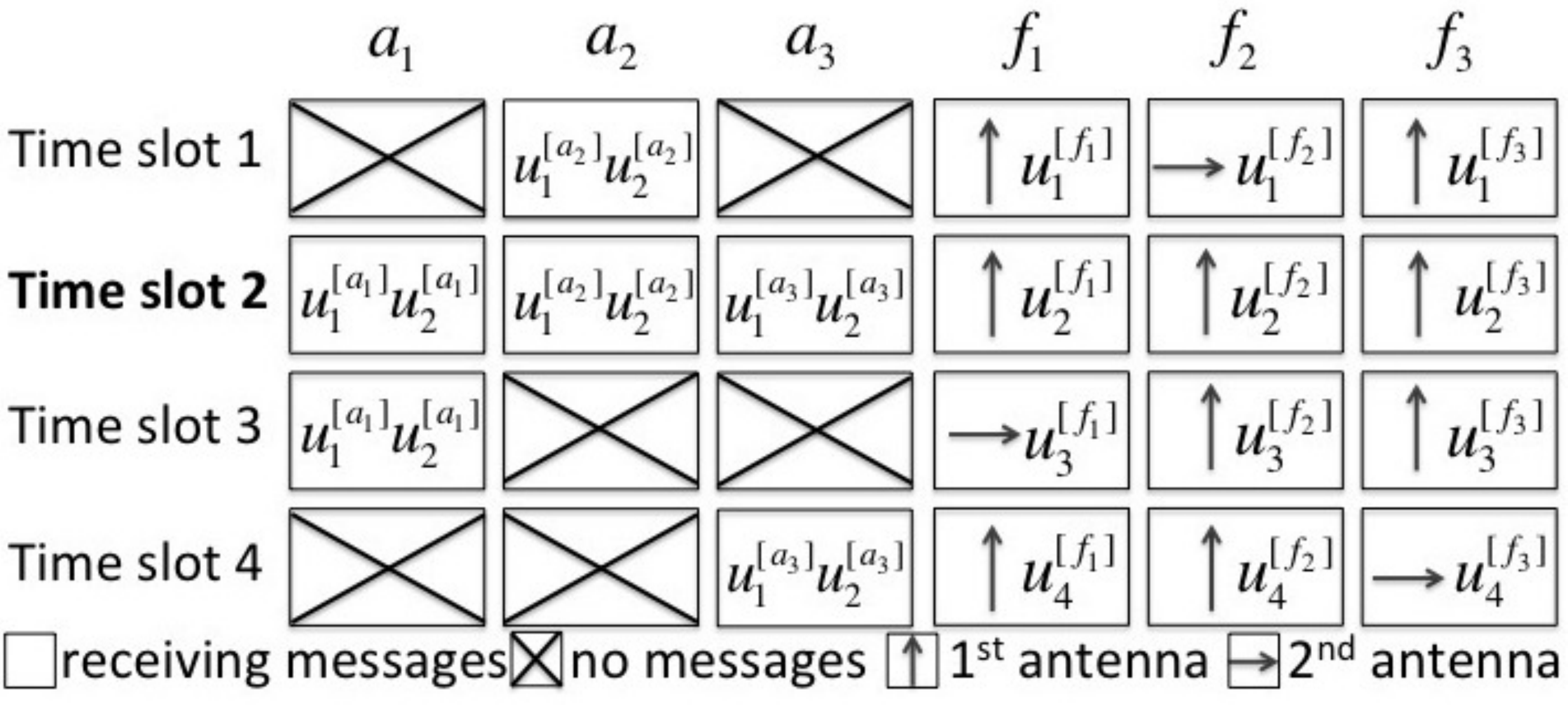}
\par\end{centering}

\caption{Beamforming in macrocell and femtocells}
\end{figure}

\begin{example}
The same model will be used as an example in this paper: For $K=3$
users in the macrocell with $N=2$ transmit/receive antennas and messages
for each user, $M_{r}=1$ receive and $N=2$ transmit antennas and
\emph{$\mathcal{M}=4$} messages sent in each one of the $K=3$ femtocells,
and $T=4$ time slots, the beamforming matrices, as shown in Figure
2, are given by:
\end{example}
\begin{center}
$\mathbf{V}^{[a_{1}]}=\frac{a}{\sqrt{N}}(\mathbf{v}^{[a_{1}]}\otimes\mathbf{I}_{2})=\frac{a}{\sqrt{2}}\bigl[\begin{array}{cccc}
0 & c & \sqrt{1-c^{2}} & 0\end{array}\bigr]^{T}\otimes\mathbf{I}_{2}$ 
\par\end{center}

\begin{center}
$\mathbf{V}^{[a_{2}]}=\frac{a}{\sqrt{N}}(\mathbf{v}^{[a_{2}]}\otimes\mathbf{I}_{2})=\frac{a}{\sqrt{2}}\bigl[\begin{array}{cccc}
\sqrt{1-c^{2}} & c & 0 & 0\end{array}\bigr]^{T}\otimes\mathbf{I}_{2}$ 
\par\end{center}

\begin{center}
$\mathbf{V}^{[a_{3}]}=\frac{a}{\sqrt{N}}(\mathbf{v}^{[a_{3}]}\otimes\mathbf{I}_{2})=\frac{a}{\sqrt{2}}\bigl[\begin{array}{cccc}
0 & c & 0 & \sqrt{1-c^{2}}\end{array}\bigr]^{T}\otimes\mathbf{I}_{2}$ 
\par\end{center}

\subsubsection{Femtocells}

At each femtocell, the \emph{($(M_{r}T)\times1$)} signal at receiver
$f_{k}$, for the supersymbol, is given by:
\begin{equation}
\mathbf{y}^{[f_{k}]}=\mathbf{H}^{[f_{k}]}\mathbf{X}{}_{f_{k}}+\mathbf{Z}^{[f_{k}]},
\end{equation}
where $\mathbf{H}^{[f_{k}]}\in\mathcal{C}^{M_{r}T\times NT}$ is the
channel transfer matrix from \emph{$T_{xf}$ }to$f_{k}$, and is given
by ${\color{black}\mathbf{H}^{[f_{k}]}=\mathbf{I}_{T}\otimes\mathbf{h}^{[f_{k}]}}$
where $\mathbf{h}^{[f_{k}]}\in\mathcal{\mathcal{C}}^{M_{r}\times N}$
is the channel for one time slot, and $\mathbf{Z}^{[f_{k}]}\sim(0,\sigma_{n}^{2}\mathbf{I}_{M_{r}T})$
denotes the Additive White Gaussian Noise (AWGN) vector.

In each femtocell, the ($\mathcal{M}\times1$) data stream vector
of each user\emph{ $f_{k}$} is given by $\mathbf{U}^{[f_{k}]}$.
The \emph{$((NT)\times\mathcal{M})$} beamforming matrix \textbf{$\mathbf{V}^{[f_{k}]}$}
is given by\textbf{:
\begin{equation}
\mathbf{V}^{[f_{k}]}=\frac{b}{\sqrt{N}}\left(\sum_{i=1}^{T}\mathbf{\mathbf{\gamma}}_{i}^{[f_{k}]^{T}}\otimes\mathbf{r}_{i}\mathbf{q}_{i}\right)
\end{equation}
}where $b\in\mathrm{\mathfrak{\boldsymbol{\mathfrak{\mathcal{\mathbb{R}}}}}}$
is a constant determined by power considerations (see (8)), and $\mathbf{v}_{1}^{[f_{k}]}=\sum_{i=1}^{T-1}\mathbf{\gamma}_{i}^{[f_{k}]}$
and $\mathbf{v}_{2}^{[f_{k}]}=\mathbf{\gamma}_{T}^{[f_{k}]}$ are\emph{
(}$1\times T$) unit vectors with entries equal to 1 and 0, with $\mathbf{v}_{2}^{[f_{k}]}$
having only its $t$th entry ($t$ denoting the time slot that $a_{k}$
receives no interference) equal to 1, such that $\sum_{j=1}^{2}\mathbf{v}_{j}^{[f_{k}]}=\begin{bmatrix}1 & 1 & ... & 1\end{bmatrix}$.
Also, for $i=1,...,T-1$, we set $\mathbf{r}_{i}$ equal to the first
$M_{r}$ columns of $\mathbf{I}_{N}$ with $\mathbf{e}_{1}$ equal
to the sum of the columns of $\mathbf{r}_{i}$, and $\mathbf{e}_{2}=\mathbf{r}_{T}$
equal to the last column of $\mathbf{I}_{N}$. Furthermore, for $i=1,...,T-1$,
$\mathbf{q}_{i}$ is equal to the submatrix of $\mathbf{I}_{\mathcal{M}}$
consisting of rows $(M_{r}(i-1)+1,M_{r}i),$ and $\mathbf{q}_{T}$
is equal to the submatrix of $\mathbf{I}_{\mathcal{M}}$ consisting
of row $\mathcal{M}$. The $t$th component of $\mathbf{\gamma}_{i}^{[f_{k}]}$
being 1 means that in the $k$th femtocell, the antennas determined
by $\mathbf{r}_{i}$ are in use at time $t$, and the messages determined
by $\mathbf{q}_{i}$ are transmitted. Finally, the ($(NT)\times1$)
vector, transmitted by \emph{$T_{xK}$} is given by:
\begin{equation}
\mathbf{X}{}_{f_{k}}=\mathbf{V}^{[f_{k}]}\mathbf{U}^{[f_{k}]}
\end{equation}

The total transmit power is given by the power constraint: 
\begin{equation}
P_{\mathrm{femtocell}}=\mathbb{E}[\mathrm{tr}(\mathbf{X}_{f_{k}}\mathbf{X}_{f_{k}}^{T})]=\mathcal{M}^{2}\frac{b^{2}}{N}
\end{equation}

\begin{example}
For our example-model, the beamforming matrix for user \emph{$f_{1}$},
as depicted in Figure 2, is given by:
\end{example}
\begin{center}
$\mathbf{V}^{[f_{1}]}=\frac{b}{\sqrt{2}}\left(\sum_{i=1}^{4}\mathbf{\mathbf{\gamma}}_{i}^{[f_{1}]^{T}}\otimes\mathbf{r}_{i}\mathbf{q}_{i}\right)$ 
\par\end{center}

\begin{center}
with $\sum_{i=1}^{3}\mathbf{\gamma}_{i}^{[f_{1}]}=\mathbf{v}_{1}^{[f_{1}]}=[\begin{array}{cccc}
{\normalcolor 1} & 1 & {\normalcolor 0} & 1\end{array}]$ , $\mathbf{\gamma}_{4}^{[f_{1}]}=\mathbf{v}_{2}^{[f_{1}]}=[\begin{array}{cccc}
{\color{black}0} & {\color{black}0} & 1 & {\color{black}0}\end{array}]$ 
\par\end{center}

\begin{center}
for $i=1,2,3:$ $\mathbf{r}_{i}=\begin{bmatrix}1 & 0\end{bmatrix}^{T}$with
$\mathbf{e}_{1}=\begin{bmatrix}1 & 0\end{bmatrix}^{T}$, $\mathbf{r}_{4}=\mathbf{e}_{2}=\begin{bmatrix}0 & 1\end{bmatrix}^{T}$,
\par\end{center}

\begin{center}
 $\mathbf{q}_{i}$ the \emph{i}th unit basis vector
\par\end{center}

\subsection{Projection \& Effective Channel Matrix}

\begin{figure}
\begin{centering}
\includegraphics[width=0.45\columnwidth]{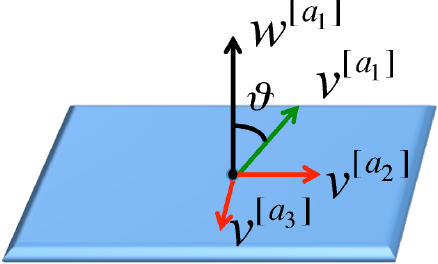}
\par\end{centering}

\caption{Schematic plot of beamforming vectors: $\mathbf{w}_{s}^{[a_{1}]}$
is orthogonal to $\mathbf{v}^{[a_{2}]}$ ,$\mathbf{v}^{[a_{3}]}$
.}
\end{figure}

\subsubsection{Macrocell}

In the macrocell, in order remove inter- and intra-cell interference,
the received signal should be projected to a subspace orthogonal to
the subspace that interference lies in. The rows of the ($N\times NT$)
projection matrix $\mathbf{P}^{[a_{k}]}$ form an orthonormal basis
of this subspace:
\begin{equation}
\mathbf{P}^{[a_{k}]}={\displaystyle \sum_{s=1}^{2}}\left(\mathbf{w}_{s}^{[a_{k}]}\otimes\mathbf{D}_{s}^{[a_{k}]}\mathbf{\widetilde{h}}^{[f_{k}a_{k}]}\right),
\end{equation}
where
\begin{enumerate}
\item for all s, the ($1\times T$) $\mathbf{w}_{s}^{[a_{k}]}$ is a unit
vector orthogonal to $\mathbf{v}^{[a_{i}]}$ for $i\neq k$, as shown
in Figure 3, 
\item \textbf{$\mathbf{w}_{s}^{[a_{k}]}$ }has coefficients equal to zero
on the non-zero values of $\mathbf{\mathbf{\gamma}}_{i}^{[f_{k}]^{T}}$
for $s=1$ and $i=T$, and for $s=2$ and $i=1...T-1$,
\item \textbf{$\mathbf{D}_{1}^{[a_{k}]}=\mathrm{diag}(\mathbf{e}_{2})$
}and \textbf{$\mathbf{D}_{2}^{[a_{k}]}=\mathrm{diag}(\mathbf{e}_{1})$, }
\item \textbf{$\mathbf{\widetilde{h}}^{[f_{k}a_{k}]}$ }is an $(N\times N)$
matrix, whose rows are unit vectors, with the $N$th row orthogonal
to all the columns of $\left(\mathbf{h}^{[f_{k}a_{k}]}\mathbf{r}_{i}\right)$
for $i=1...T-1$, and the remaining $(N-1)$ rows orthogonal to $\left(\mathbf{h}^{[f_{k}a_{k}]}\mathbf{\mathbf{r}}_{T}\right)$.\end{enumerate}
\begin{example}
For the toy-model, setting $A=\sqrt{(h_{22}^{[f_{1}a_{1}]2}+h_{12}^{[f_{1}a_{1}]2})}$
and $B=\sqrt{(h_{21}^{[f_{1}a_{1}]2}+h_{11}^{[f_{1}a_{1}]2})}$, $P^{[a_{1}]}$
is given by:\textbf{
\[
\mathbf{P}^{[a_{1}]}={\displaystyle \sum_{s=1}^{2}}\left(\mathbf{w}_{s}^{[a_{1}]}\otimes\mathbf{D}_{s}^{[a_{1}]}\mathbf{\widetilde{h}}^{[f_{1}a_{1}]}\right)=
\]
\[
\left(\begin{bmatrix}\frac{c}{\sqrt{1+c^{2}}} & \frac{-\sqrt{1-c^{2}}}{\sqrt{1+c^{2}}} & 0 & \frac{c}{\sqrt{1+c^{2}}}\end{bmatrix}\otimes\begin{bmatrix}0 & 0\\
\frac{h_{21}^{[f_{1}a_{1}]}}{B} & \frac{-h_{11}^{[f_{1}a_{1}]}}{B}
\end{bmatrix}\right)
\]
\[
+\left(\begin{bmatrix}0 & 0 & 1 & 0\end{bmatrix}\otimes\begin{bmatrix}\frac{h_{22}^{[f_{1}a_{1}]}}{A} & \frac{-h_{12}^{[f_{1}a_{1}]}}{A}\\
0 & 0
\end{bmatrix}\right),
\]
} where 
\end{example}
\begin{center}
$\mathbf{w}_{1}^{[a_{1}]}=\begin{bmatrix}\frac{c}{\sqrt{1+c^{2}}} & \frac{-\sqrt{1-c^{2}}}{\sqrt{1+c^{2}}} & 0 & \frac{c}{\sqrt{1+c^{2}}}\end{bmatrix}$,
$\mathbf{w}_{2}^{[a_{1}]}=\begin{bmatrix}0 & 0 & 1 & 0\end{bmatrix}$,
\par\end{center}

\begin{center}
$\mathbf{D}_{1}^{[a_{k}]}=\mathrm{diag}(\begin{bmatrix}0 & 1\end{bmatrix}^{T})$,
$\mathbf{D}_{2}^{[a_{k}]}=\mathrm{diag}(\begin{bmatrix}1 & 0\end{bmatrix}^{T})$,
\par\end{center}

\begin{center}
$\widetilde{\mathbf{h}}^{[f_{1}a_{1}]}=\begin{bmatrix}\frac{1}{A} & 0\\
0 & \frac{1}{B}
\end{bmatrix}\begin{bmatrix}h_{22}^{[f_{1}a_{1}]} & -h_{12}^{[f_{1}a_{1}]}\\
h_{21}^{[f_{1}a_{1}]} & -h_{11}^{[f_{1}a_{1}]}
\end{bmatrix}$
\par\end{center}

\emph{Theorem 1.}\textbf{ }Multiplying the received signal by projection
matrix \textbf{$\mathbf{P}^{[a_{k}]}$}: 
\begin{equation}
\mathbf{\widetilde{y}}^{[a_{k}]}=\mathcal{\mathbf{P}}^{[a_{k}]}\mathbf{y}^{[a_{k}]}
\end{equation}
gives an effective channel:
\begin{equation}
\mathbf{\widetilde{y}}^{[a_{k}]}=\mathcal{\mathbf{\mathcal{H}}}^{[a_{k}]}\mathbf{U}^{[a_{k}]}+\mathbf{\widetilde{Z}}^{[a_{k}]},
\end{equation}
where\textbf{ 
\begin{equation}
\mathcal{\mathbf{\mathcal{H}}}^{[a_{k}]}=\frac{a}{\sqrt{N}}\mathcal{D}^{[a_{k}]}\mathbf{\widetilde{h}}^{[f_{k}a_{k}]}\mathbf{h}^{[a_{k}]},
\end{equation}
}with diagonal matrix\textbf{ }
\begin{equation}
\mathcal{D}^{[a_{k}]}={\displaystyle \sum_{s=1}^{2}}\mathbf{w}_{s}^{[a_{k}]}\mathbf{v}^{[a_{k}]}\mathbf{D}_{s}^{[a_{k}]}=\mathrm{diag}\left(\mathbf{w}_{s}^{[a_{k}]}\mathbf{v}^{[a_{k}]}\right)
\end{equation}
and $\mathbf{\widetilde{Z}}^{[a_{k}]}=\mathcal{\mathbf{P}}^{[a_{k}]}\mathbf{Z}^{[a_{k}]}$
remains white noise with the same variance (since $\mathbf{w}_{s}^{[a_{k}]}$
is a unit vector).
\begin{IEEEproof}
We show that $P^{[a_{k}]}$ removes intra- and inter- cell interference
at the \emph{k}th receiver. Substituting, (1) and (3) in (10), we
can consider the coefficients of $\mathbf{U}^{[a_{i}]}$ and $\mathbf{U}^{[f_{k}]}$
separately.\textbf{ }For $i\neq k$, using $\left(A\otimes B\right)\left(C\otimes D\right)=\left(AC\right)\otimes\left(BD\right)$,
for intra-cell interference, coefficient of $\mathbf{U}^{[a_{i}]}$
becomes:
\begin{align*}
\mathbf{P}^{[a_{k}]}\mathbf{H}^{[a_{k}]}\mathbf{V}^{[a_{i}]} & =\frac{a}{\sqrt{N}}{\displaystyle \sum_{s=1}^{2}}\left(\mathbf{w}_{s}^{[a_{k}]}\otimes\mathbf{D}_{s}^{[a_{k}]}\mathbf{\widetilde{h}}^{[f_{k}a_{k}]}\right)\\
 & \times(\mathbf{I}_{T}\otimes\mathbf{h}^{[a_{k}]})(\mathbf{v}^{[a_{i}]}\otimes\mathbf{I}_{N})
\end{align*}
\begin{equation}
=\frac{a}{\sqrt{N}}{\displaystyle \sum_{s=1}^{2}}\left(\mathbf{w}_{s}^{[a_{k}]}\mathbf{v}^{[a_{i}]}\right)\otimes\left(\mathbf{D}_{s}^{[a_{k}]}\mathbf{\widetilde{h}}^{[f_{k}a_{k}]}\mathbf{h}^{[a_{k}]}\right),
\end{equation}
where by definition, for all s, $\mathbf{w}_{s}^{[a_{k}]}\mathbf{v}^{[a_{i}]}=0$,
i.e. $\mathbf{w}_{s}^{[a_{k}]}$ is orthogonal to $\mathbf{v}^{[a_{i}]}$
if $i\neq k$. For $i=k$ the remaining term is (13). For inter-cell
interference, coefficient of $\mathbf{U}^{[f_{k}]}$:
\begin{align*}
\mathbf{P}^{[a_{k}]}\mathbf{H}^{[f_{k}a_{k}]}\mathbf{V}^{[f_{k}]} & =\frac{b}{\sqrt{M_{t}}}{\displaystyle \sum_{s=1}^{2}}\left(\mathbf{w}_{s}^{[a_{k}]}\otimes\mathbf{D}_{s}^{[a_{k}]}\mathbf{\widetilde{h}}^{[f_{k}a_{k}]}\right)\\
 & \times\left(\mathbf{I}_{T}\otimes\mathbf{h}^{[f_{k}a_{k}]}\right)\left(\sum_{i=1}^{T}\mathbf{\mathbf{\gamma}}_{i}^{[f_{k}]^{T}}\otimes\mathbf{r}_{i}\mathbf{q}_{i}\right)
\end{align*}
\begin{equation}
=\frac{b}{\sqrt{M_{t}}}{\displaystyle \sum_{s=1}^{2}{\displaystyle \sum_{i=1}^{T}\mathbf{w}_{s}^{[a_{k}]}}}\mathbf{\gamma}_{i}^{[f_{k}]^{T}}\otimes\mathbf{D}_{s}^{[a_{k}]}\mathbf{\widetilde{h}}^{[f_{k}a_{k}]}\mathbf{h}^{[f_{k}a_{k}]}\mathbf{r}_{i}\mathbf{q}_{i},
\end{equation}
where for \textbf{$\mathbf{s=1}$:} if $i=T$, the $(\mathbf{w}_{s}^{[a_{k}]}\mathbf{\gamma}_{i}^{[f_{k}]^{T}})=0$
and if $i=1,...,(T-1)$, the $(\mathbf{D}_{s}^{[a_{k}]}\mathbf{\widetilde{h}}^{[f_{k}a_{k}]}\mathbf{h}^{[f_{k}a_{k}]}\mathbf{r}_{i})=0$.
Premultiplying by $\mathbf{D}_{s}^{[a_{k}]}$ selects a row of $\mathbf{\widetilde{h}}^{[f_{k}a_{k}]}$
and postmultiplying by $\mathbf{r}_{i}$ selects a column of $\mathbf{\widetilde{h}}^{[f_{k}a_{k}]}$,
with the resulting row and column being orthogonal by 4).

For\textbf{ $\mathbf{s=2}$:} if $i=1,...,(T-1)$, the $(\mathbf{w}_{s}^{[a_{k}]}\mathbf{\gamma}_{i}^{[f_{k}]^{T}})=0$
and if $i=T$, the $(\mathbf{D}_{s}^{[a_{k}]}\mathbf{\widetilde{h}}^{[f_{k}a_{k}]}\mathbf{h}^{[f_{k}a_{k}]}\mathbf{r}_{i})=0$. 
\end{IEEEproof}

\subsubsection{Femtocell}

The effective channel matrix $\mathcal{\mathbf{\mathcal{H}}}^{[f_{k}]}$
is given by:
\begin{equation}
\mathcal{\mathbf{\mathcal{H}}}^{[f_{k}]}=\frac{b}{\sqrt{N}}\left(\mathbf{I}_{T}\otimes\mathbf{h}^{[f_{k}]}\right)\left(\sum_{i=1}^{T}\mathbf{\mathbf{\gamma}}_{i}^{[f_{k}]^{T}}\otimes\mathbf{r}_{i}\mathbf{q}_{i}\right),
\end{equation}
and the final post-processed signal at receiver $f_{k}$ becomes:
\begin{equation}
\mathbf{\widetilde{y}}^{[f_{k}]}=\mathcal{\mathbf{\mathcal{H}}}^{[f_{k}]}\mathbf{U}^{[f_{k}]}+\mathbf{\widetilde{Z}}^{[f_{k}]},
\end{equation}
where $\mathbf{\widetilde{Z}}^{[f_{k}]}$ remains white noise with
the same variance.

\section{Degrees of Freedom}

\emph{Theorem 2: }In the heterogeneous network, counting messages,
$DoF_{\mathrm{macrocell}}=\frac{KN}{K+1}$ and $DoF_{\mathrm{femtocell}}=\frac{KM_{r}+1}{K+1}$,
and thus the total DoF that can be achieved are given by:
\begin{equation}
DoF_{\mathrm{total}}=\frac{K(N+KM_{r}+1)}{K+1}
\end{equation}

\subsection{BIA vs. TDMA}

In order to further understand the advantage, in DoF, of the BIA scheme
proposed, Table 1 summarizes the DoF that can be achieved by BIA and
TDMA. The total DoF gain achieved by BIA is given by $DoF_{BIA}-DoF_{TDMA}=\frac{K-N+M_{r}}{K+1}\underset{(M_{r}=N-1)}{=}\frac{K-1}{K+1}$. 

\begin{center}
\begin{tabular}{|c|c|c|c|}
\hline 
\textbf{Scheme} & \textbf{Macrocell} & \textbf{K Femtocells} & \textbf{Total Network}\tabularnewline
\hline 
\hline 
\textbf{BIA} & $\frac{KN}{K+1}$ & $\frac{K(KM_{r}+1)}{K+1}$ & $\frac{K(N+KM_{r}+1)}{K+1}$\tabularnewline
\hline 
\textbf{TDMA} & $N$ & $M_{r}(K-1)$ & $N+M_{r}(K-1)$\tabularnewline
\hline 
\end{tabular}
\par\end{center}

\begin{center}
Table 1: DoF of BIA and TDMA
\par\end{center}

Moreover, the benefit of the employment of the BIA scheme in heterogeneous
networks is related to the number of receive antennas in the macrocell
and femtocells. Based on our research, compared to TDMA, as the number
of receive antennas in the macrocell increases, the benefit we get
from BIA decreases. Finally, as the number of receive antennas in
each femtocell increases, the benefit we get from BIA remains almost
the same.

\section{Achievable Rate }

\subsection{Macrocell}

Since there is no CSIT, the total rate for each user in the macrocell,
for ONE time slot and setting $\mathcal{K}^{[a_{k}]}=\widetilde{\mathbf{h}}^{[f_{k}a_{k}]}\mathbf{h}^{[a_{k}]}$,
is given by:
\begin{equation}
R^{[a_{k}]}=\frac{1}{T}\mathrm{\mathtt{\mathbb{E}}}\left[\log\det\left(\mathbf{I}_{N}+\frac{P_{\mathrm{macrocell}}}{KN^{2}\sigma_{n}^{2}}\mathcal{D}^{[a_{k}]}\mathcal{K}^{[a_{k}]}\mathcal{K}^{[a_{k}]^{*}}\mathcal{D}^{[a_{k}]^{*}}\right)\right]
\end{equation}

For any channel realisation, in the high SNR limit, the rate is maximised
by maximising the value of 
\begin{equation}
\mathrm{det}\mathcal{D}^{[a_{k}]}=\prod_{s=1}^{2}\left(\mathbf{w}_{s}^{[a_{k}]}\mathbf{v}^{[a_{k}]}\right)
\end{equation}
For our example, see Figure 3, with $K=3$ users, (20) is maximised
for $c=0.5299,$ since the values of $N$ and channel transfer matrices
are fixed for a given channel realisation.

\subsection{Femtocells}

Since there is no CSIT, the rate for each femtocell user, for ONE
time slot, is given by:
\begin{equation}
R^{[f_{k}]}=\frac{1}{T}\mathcal{\mathbb{E}}\left[\log\det\left(\mathbf{I}_{\mathcal{M}}+\frac{P_{\mathrm{femtocell}}}{\sigma_{n}^{2}}\frac{1}{\mathcal{M}^{2}}\mathcal{\widetilde{H}}^{[f_{k}]}\mathcal{\widetilde{H}}^{[f_{k}]^{*}}\right)\right],
\end{equation}
where
\begin{equation}
\mathcal{\mathbf{\mathcal{\widetilde{H}}}}^{[f_{k}]}=\frac{\sqrt{N}}{b}\mathcal{\mathbf{\mathcal{H}}}^{[f_{k}]}
\end{equation}

\section{Overview of results}

Most of our simulations were based on the toy-model already described.
The statistical model chosen was i.i.d. Rayleigh and our input symbols
were QPSK modulated. Finally, Zero-Forcing (ZF) detection was performed
in the decoding stage. Typical values of $a$ and $b$ used in a real
system are $a=\sqrt{40}$ and $b=\sqrt{5}$.

\subsection{Bit Error Rate (BER) Performance}

In order to investigate the BER performance of our toy-model, the
effect of varying the values of constants $a$ and $c$ from the beamforming
vectors for macrocell users and $b$ from the beamforming vectors
for femtocell users, was investigated. Firstly, the BER performance
of the network was investigated for different values of $a$ and $b$.
As $a$ and $b$ ``control'' the power with which messages are transmitted,
when they are varied, effectively the total transmit power of the
network changes as well. For instance, Figure 4 shows how the BER
performances of the macrocell and femtocells are affected when we
vary coefficients $a$ and $b$. Finally, as discussed in Section
IV an optimal value for $c$ can be found, which for our toy-model
is $c=0.5299$. Figure 5 depicts how the BER performance of the macrocell
is affected as $c$ changes. 

\begin{figure}
\begin{centering}
\includegraphics[width=0.8\columnwidth]{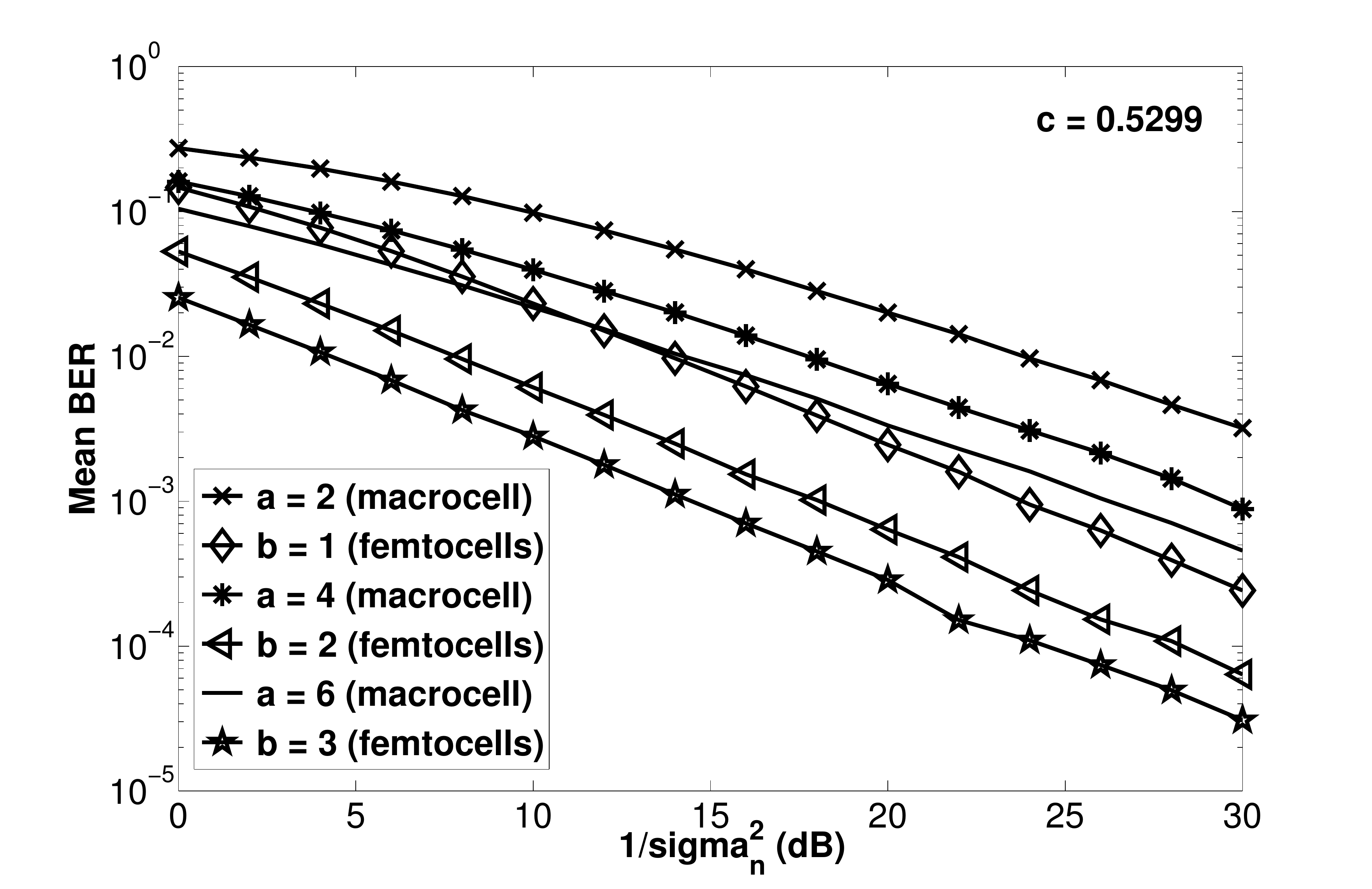}
\caption{BER in macrocell and femtocells for different values of $a$ and $b$.
As $a$ and $b$ increase BER performances in macrocell and femtocells
improve.}
\par \end{centering}
\end{figure}

\begin{figure}
\begin{centering}
\includegraphics[width=0.8\columnwidth]{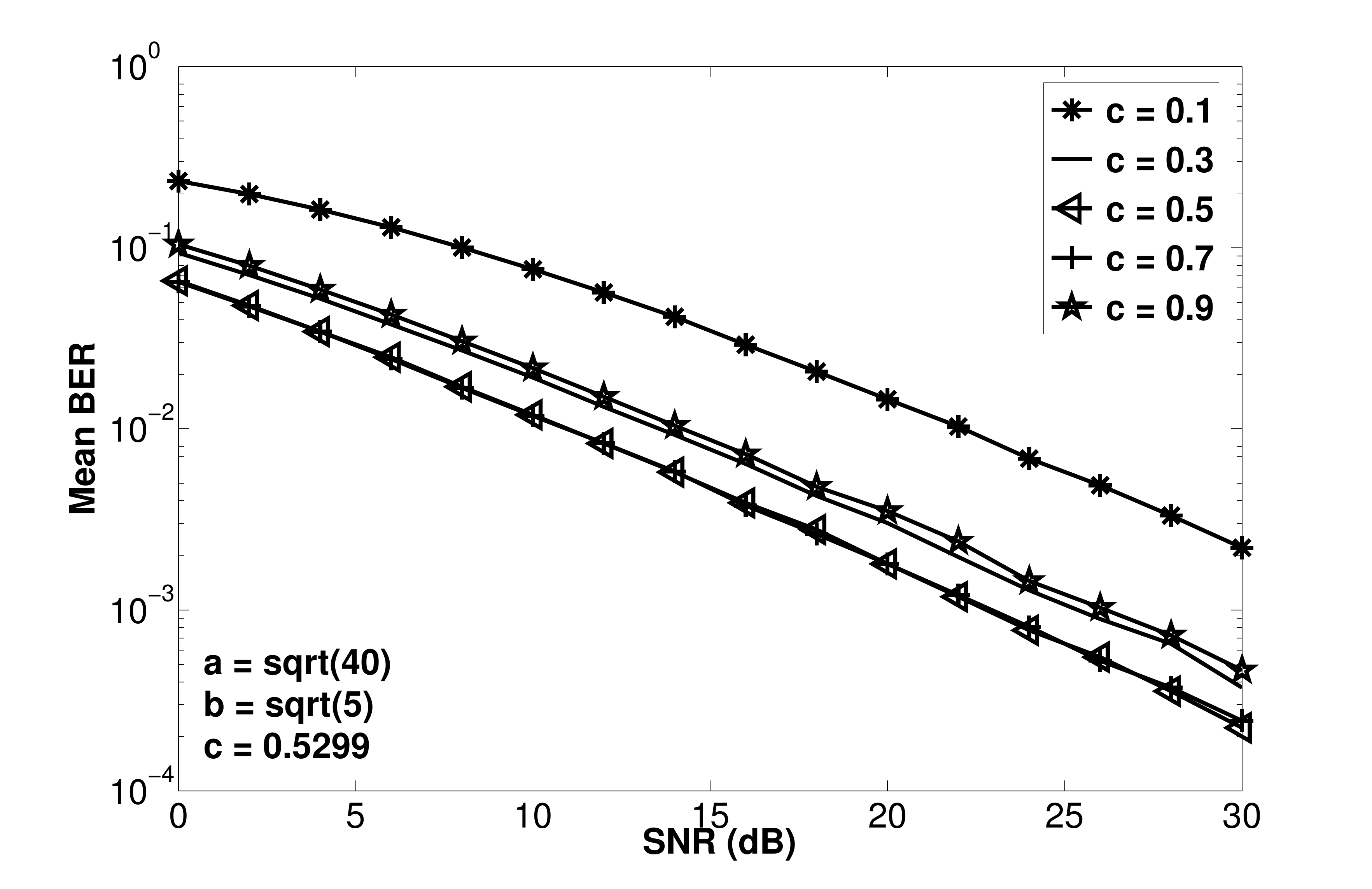}
\par\end{centering}
\caption{BER in the macrocell for different values of $c$. BER performance
in the macrocell is improved for values of $c$ close to 0.5299.}
\end{figure}

\begin{figure}
\begin{centering}
\includegraphics[width=0.8\columnwidth]{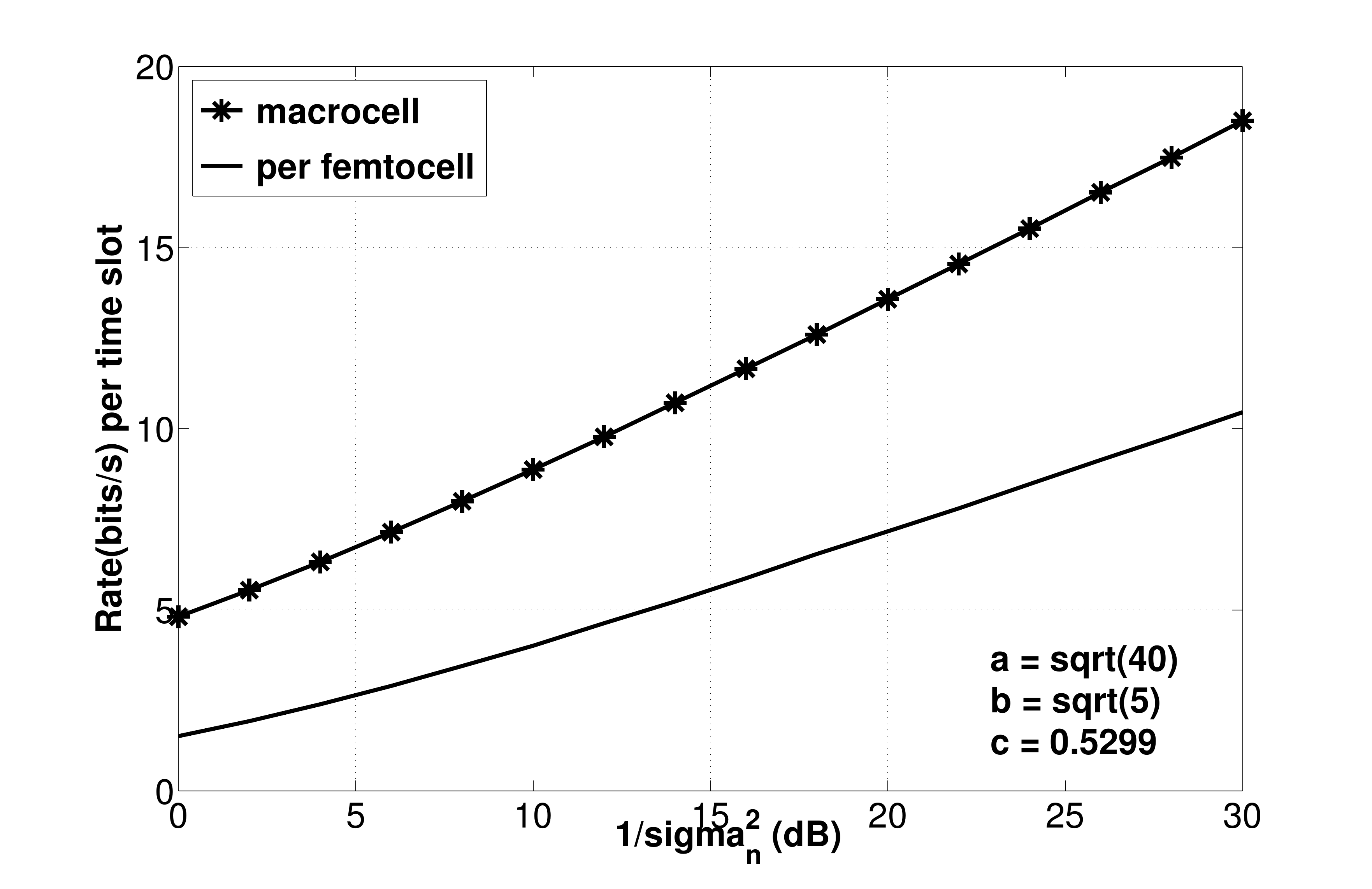}
\par\end{centering}

\caption{Rate of heterogeneous network with respect to the noise variance.
The rate in the macrocell is higher than the rate per femtocell.}
\end{figure}

\begin{figure}
\begin{centering}
\includegraphics[width=0.8\columnwidth]{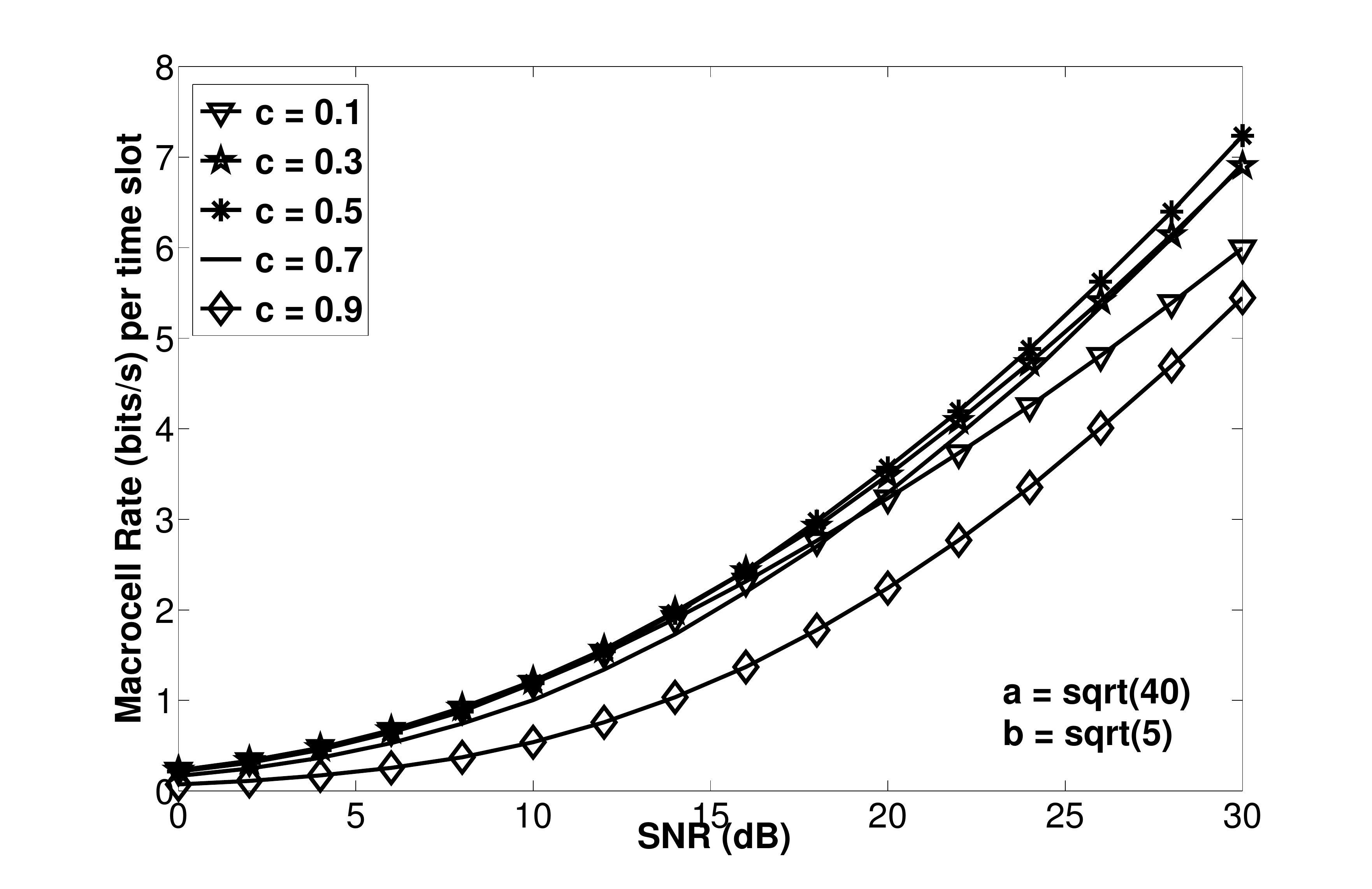}
\par\end{centering}

\caption{Rate in the macrocell for different values of $c$. The rate in the
macrocell is optimised for values of $c$ very close to 0.5299.}
\end{figure}

\subsection{Sum Rate Performance}

As discussed in section IV, the value of $c$ has a key-role in the
sum rate performance of the macrocell. In Figure 6 the rate of the
heterogeneous network is depicted and in Figure 7, it can be observed
how the sum rate in the macrocell changes with $c$, achieving its
best performance for values of $c$ close to $0.5299$.

\section{Summary}

Overall, this paper introduces how the BIA scheme can be applied into
heterogeneous networks. Considering the fact that no CSIT is required,
the DoF that can be achieved were discussed, which are the same with
the IA scheme requiring perfect CSIT. Moreover, the BIA model was
investigated from the perspective of equal power allocation, and how
that can affect the optimal performance of the system. In that context,
the important role of $c$, in the performance of the network, suggests
that there is ground for further research on optimising the network
performance. Finally, the description of the model in a Kronecker
product representation provides a different insight on how the BIA
scheme works.

\section{Acknowledgements}

This work was supported by NEC; the Engineering and Physical Sciences
Research Council {[}EP/I028153/1{]}; and the University of Bristol.
The authors thank Simon Fletcher \& Patricia Wells for useful discussions.

\end{document}